\documentclass[twocolumn,aps,amsfonts,amssymb,amsmath,floats,floatfix,twocolumn,superscriptaddress,unsortedaddress]{revtex4}
\usepackage{graphics}
\usepackage{epsfig,graphicx}

\begin{document}

\title{Electronic superlattice revealed by resonant scattering \\
from random impurities in Sr$_3$Ru$_2$O$_7$}

\author{M.A. Hossain}
\thanks{Correspondence to: mahossain@lbl.gov}
\affiliation{Department of Physics {\rm {\&}} Astronomy, University of British Columbia, Vancouver, British Columbia V6T\,1Z1, Canada}
\affiliation{Advanced Light Source, Lawrence Berkeley National Laboratory, Berkeley, California 94720, USA}
\author{I. Zegkinoglou}
\affiliation{Max-Planck-Institut f\"{u}r Festk\"{o}rperforschung, Heisenbergstra\ss e 1, 70569 Stuttgart, Germany}
\author{Y.-D. Chuang}
\affiliation{Advanced Light Source, Lawrence Berkeley National Laboratory, Berkeley, California 94720, USA}
\author{J. Geck}
\affiliation{Department of Physics {\rm {\&}} Astronomy, University of British Columbia, Vancouver, British Columbia V6T\,1Z1, Canada}
\author{B. Bohnenbuck}
\affiliation{Max-Planck-Institut f\"{u}r Festk\"{o}rperforschung, Heisenbergstra\ss e 1, 70569 Stuttgart, Germany}
\author{A.G. Cruz Gonzalez}
\affiliation{Advanced Light Source, Lawrence Berkeley National Laboratory, Berkeley, California 94720, USA}
\author{H.-H. Wu}
\affiliation{II. Physikalisches Institut, Universit\"{a}t zu K\"{o}ln, Z\"{u}lpicher Stra\ss e 77, 50937 K\"{o}ln, Germany}
\author{C. Sch\"{u}\ss ler-Langeheine}
\affiliation{II. Physikalisches Institut, Universit\"{a}t zu K\"{o}ln, Z\"{u}lpicher Stra\ss e 77, 50937 K\"{o}ln, Germany}
\author{D.G. Hawthorn}
\affiliation{Department of Physics {\rm {\&}} Astronomy, University of British Columbia, Vancouver, British Columbia V6T\,1Z1, Canada}
\author{J.D. Denlinger}
\affiliation{Advanced Light Source, Lawrence Berkeley National Laboratory, Berkeley, California 94720, USA}
\author{R. Mathieu}
\affiliation{Department of Applied Physics, University of Tokyo, Tokyo 113-8656, Japan}
\author{Y. Tokura}
\affiliation{Department of Applied Physics, University of Tokyo, Tokyo 113-8656, Japan}
\author{S. Satow}
\affiliation{Department of Advanced Materials Science, University of Tokyo, Kashiwa, Chiba 277-8581, Japan}
\author{H. Takagi}
\affiliation{Department of Advanced Materials Science, University of Tokyo, Kashiwa, Chiba 277-8581, Japan}
\author{\\Y. Yoshida}
\affiliation{National Institute of Advanced Industrial Science and Technology  (AIST), Tsukuba, 305-8568, Japan}
\author{Z. Hussain}
\affiliation{Advanced Light Source, Lawrence Berkeley National Laboratory, Berkeley, California 94720, USA}
\author{B. Keimer}
\affiliation{Max-Planck-Institut f\"{u}r Festk\"{o}rperforschung, Heisenbergstra\ss e 1, 70569 Stuttgart, Germany}
\author{G.A. Sawatzky}
\affiliation{Department of Physics {\rm {\&}} Astronomy, University of British Columbia, Vancouver, British Columbia V6T\,1Z1, Canada}
\affiliation{Quantum Matter Institute, University of British Columbia, Vancouver, British Columbia V6T 1Z4, Canada}
\author{A. Damascelli}
\thanks{Correspondence to: damascelli@physics.ubc.ca}
\affiliation{Department of Physics {\rm {\&}} Astronomy, University of British Columbia, Vancouver, British Columbia V6T\,1Z1, Canada}
\affiliation{Quantum Matter Institute, University of British Columbia, Vancouver, British Columbia V6T 1Z4, Canada}

\date{\today}

\maketitle

{\bf Resonant elastic x-ray scattering (REXS) is an exquisite element-sensitive tool for the study of subtle charge, orbital, and spin superlattice orders driven by the valence electrons, which therefore escape detection in conventional x-ray diffraction (XRD). Although the power of REXS has been demonstrated by numerous studies of complex oxides performed in the soft x-ray regime, the cross section and photon wavelength of the material-specific elemental absorption edges ultimately set the limit to the smallest superlattice amplitude and periodicity one can probe. Here we show -- with simulations and REXS on Mn-substituted Sr$_3$Ru$_2$O$_7$ -- that these limitations can be overcome by performing resonant scattering experiments at the absorption edge of a suitably-chosen, dilute impurity. This establishes that -- in analogy with impurity-based methods used in electron-spin-resonance, nuclear-magnetic resonance, and M\"ossbauer spectroscopy -- randomly distributed impurities can serve as a non-invasive, but now momentum-dependent probe, greatly extending the applicability of resonant x-ray scattering techniques.}

\begin{figure}[b!]
\centerline{\epsfig{figure=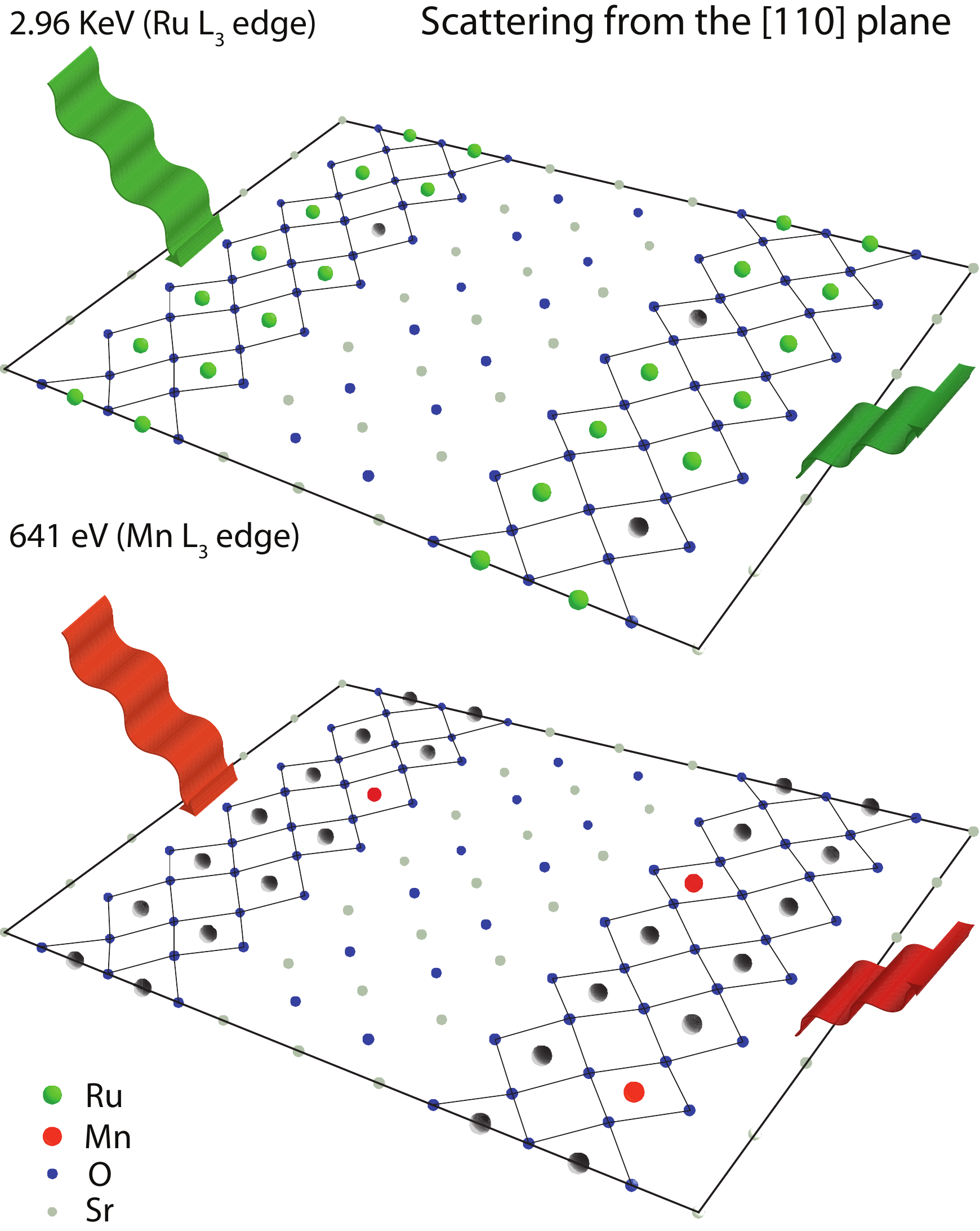,width=1.0\linewidth,clip=}} 
\caption{Scattering from Sr$_3$(Ru$_{1-x}$Mn$_x$)$_2$O$_7$ along the [110] plane; one can see the Mn-substituted RuO$_2$ bilayers, separated by SrO planes. (a) When the x-ray energy is tuned to the Ru L$_3$ edge (2.96 KeV), the scattering signal is determined by the Ru atoms (green); (b) on the converse, at the Mn L$_3$ edge (641 eV), only the Mn atoms (red), occupying $\sim\!10\%$ of the sites, contribute to the scattering signal.}\label{Fig1}
\end{figure}

Impurities are widely used as a means of controlling the physical properties of materials. A different, yet equally important use of impurities is as a probe to investigate those physical properties, in a wide range of materials and methods. For instance in biology, embedding impurities into complex molecules is a powerful way to study the local structural environment -- around each site separately -- by x-ray diffraction techniques; in particular, the inclusion of heavy atoms such as Se helps solve the phase problem in single- or multi-wavelength anomalous diffraction, so-called SAD or MAD \cite{hendrickson,larsson}. An analogous approach has also been employed in the field of complex oxides; for instance, Gd impurities with 4$f$ valence electrons have been substituted for 3$d$-Cu atoms in the high-$T_c$ superconductors YBa$_2$Cu$_3$O$_{6+x}$  and YBa$_2$Cu$_4$O$_{8}$, to probe the antiferromagnetic order incipient in the CuO$_2$ plane \cite{janossy,feher}: since the Gd moments are slaved to the Cu moments by very weak $d$-$f$ exchange interactions, Gd behaves as a truly noninvasive probe in electron-spin-resonance, nuclear-magnetic resonance, and M\"ossbauer spectroscopy experiments.

A momentum-dependent analogue of these approaches could be realized in resonant x-ray scattering techniques which, owing to the element selectivity of the resonant-absorption process and the enhanced sensitivity to excitations and superlattice modulations of the valence electrons, are becoming one of the most prominent tools in the study of incipient ordering phenomena in complex oxides. For instance, this has been demonstrated by numerous resonant elastic soft x-ray scattering (REXS) studies of spin, charge, and orbital ordering in various manganites \cite{murakami} and cuprates \cite{abbamonte}, including oxygen-ordered YBa$_2$Cu$_3$O$_{6+x}$ \cite{ghiringhelli, achkar2012,chang2012,blanco}. As we will show in the following, dilute substitutional impurities can serve as a momentum dependent probe to extend the applicability of this and other resonant x-ray scattering methods.

Conventional x-ray diffraction (XRD) is based on Thomson scattering, where the scattering amplitude is proportional to the total electron charge density of the lattice sites \cite{book}. For instance, in case of a square lattice spanned by Ru atoms, there are 44 electrons per site. If only one of the valence electrons at each site participates in forming a magnetic/charge/orbital superlattice, conventional XRD will most often not be able to detect these very weak modulations since the corresponding diffraction peaks -- with intensity proportional to the square of the electron charge density -- will be buried under the hugely intense structural ordering peaks. However, close to the absorption edge, when the x-ray energy and polarization are selectively tuned to the core-electron-to-valence-hole transitions of a specific element, scattering due to valence holes is greatly enhanced. This leads to an enhanced sensitivity to charge and spin density waves of the valence electrons and holes, making it possible to explore the most subtle spin/charge/orbital ordering phenomena in strongly correlated systems \cite{murakami, abbamonte, ghiringhelli}. This also confers to REXS its element sensitivity. By tuning the energy of the incoming light to that of a substitutional impurity, one can in principle get a resonance enhanced scattering signal only from the impurity sites. 
\begin{figure}[t!]
\centerline{\epsfig{figure=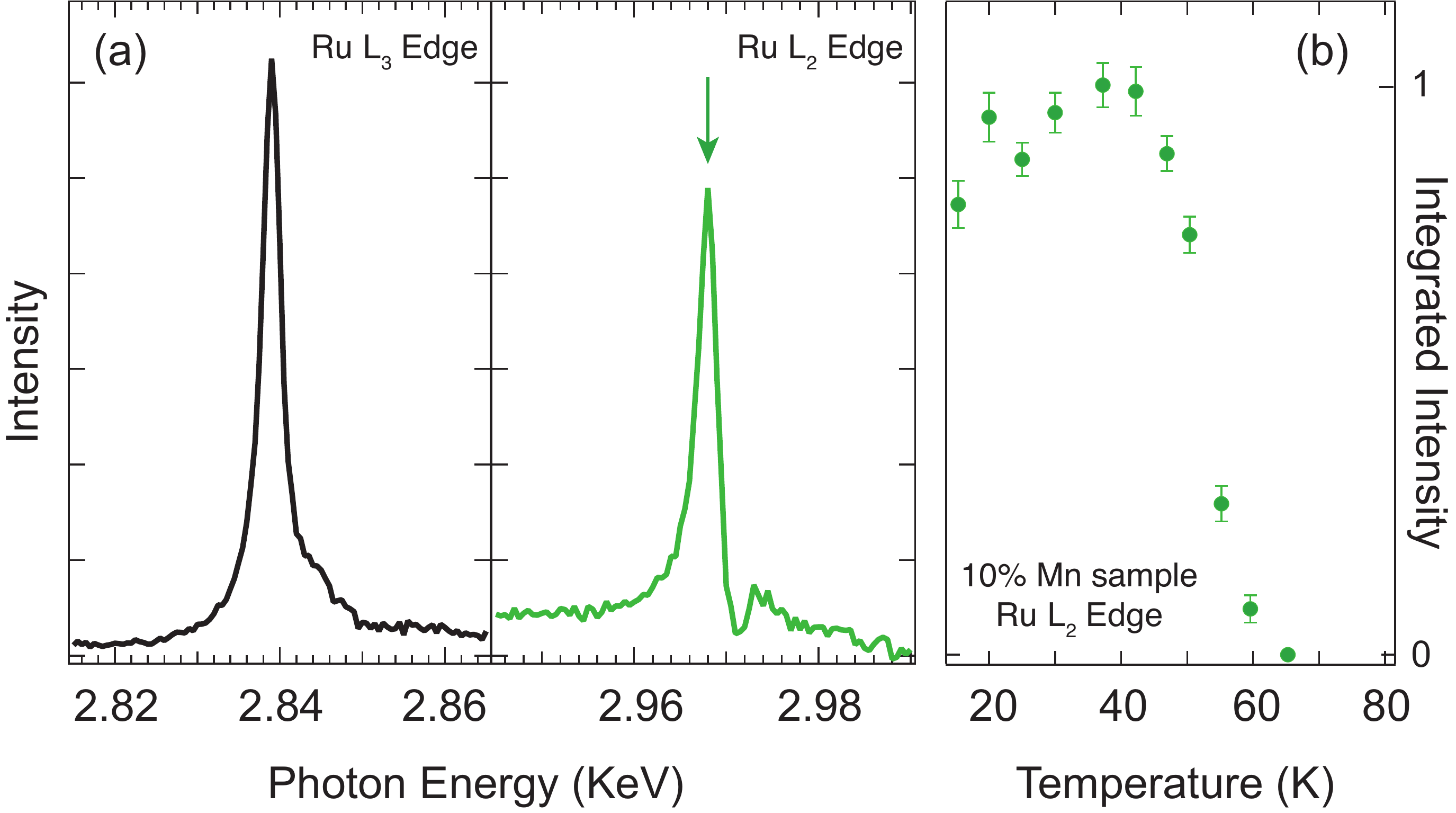,width=1.0\linewidth,clip=}} 
\caption{Ru resonance REXS results from 10\% Mn substituted  Sr$_3$(Ru$_{1-x}$Mn$_x$)$_2$O$_7$  $(x\!=\!0.1)$. (a) Energy profile of the $(\frac{1}{4},\frac{1}{4},0)$ peak across the Ru L$_{2,3}$ edge at $T\!=\!20$\,K. (b) Temperature dependence of the integrated intensity of the $(\frac{1}{4},\frac{1}{4},0)$ peak at the Ru L$_2$ edge; the energy position (2968\,eV) of the measurement is shown by the arrow in panel (a).}
\label{Fig3}
\end{figure}

Here we study Mn-substituted Sr$_3$Ru$_2$O$_7$ by REXS to demonstrate the role of Mn impurities, with $3d$ valence electrons, in revealing the underlying instabilities of the  Ru\,4$d$\,-\,O\,2$p$ valence electron of the RuO$_2$ planes of this system. The problem of measuring REXS at the impurity edge from Mn-substituted Sr$_3$Ru$_2$O$_7$ is schematically presented in Fig.\,\ref{Fig1}, which emphasizes the elemental-absorption-edge-specific contributions to the resonant scattering signal from the Ru and Mn lattice sites. To be able to observe any diffraction Bragg peaks in momentum space, we require long-range order in real space, as well as a large enough number of scattering centers. Given that the Mn impurities are randomly distributed in the system and are also dilute (Fig.\,\ref{Fig1}), one may wonder if these requirements might fundamentally prevent the detection of any Mn-edge REXS signal, even if there is some kind of order in the host RuO$_2$ plane in which the Mn impurities participate. We find that the $(\frac{1}{4},\frac{1}{4},0)$ magnetic superstructure can be detected with much higher sensitivity at the Mn than Ru absorption edge. This paves the way to the use of dilute, substitutional impurities as a `marker' of electronic ordering phenomena in other oxides and complex systems.\\

\section*{\large{Results}}

\noindent
{\bf Choice of material system.} As a study case for the electronic scattering from random impurities, let us introduce Sr$_3$(Ru$_{1-x}$Mn$_x$)$_2$O$_7$, in which Mn is being randomly substituted on the Ru site within the RuO$_2$ planes of the parent compound (see Fig.\ref{Fig1}). Pure Sr$_3$Ru$_2$O$_7$ is known as a metal on the verge of ferromagnetism \cite{ikeda}, due to the presence of strong ferromagnetic fluctuations. More recently, magnetic field tuned quantum criticality \cite{grigera} and electronic nematic fluid behavior \cite{borzi} have been proposed for this compound and associated with a metamagnetic transition. In Sr$_3$(Ru$_{1-x}$Mn$_x$)$_2$O$_7$, it has been shown that Mn impurities display an unusual crystal-field level inversion \cite{hossain_PRL}, due to the interplay between localized Mn\,3$d$ and delocalized Ru\,4$d$\,-\,O\,2$p$ valence states. Upon lowering the temperature, a Mott-type metal-insulator (MIT) phase transition driven by electronic correlations has been observed for 5\,\% Mn substitution at $T_{MIT}\!\simeq\!50$\,K, and at progressively higher $T_{MIT}$ upon increasing the Mn concentration \cite{mathieu,Hu_PRB,hossain_PRB}. Following the onset of local antiferromagnetic (AF) correlations at $T_{MIT}$, for Mn concentrations $x\!=\!0.025\!-\!0.2$ an unusual long-range AF superstructure emerges at $T_{order}\!<\!T_{MIT}$ \cite{Hu_PRB,hossain_PRB,Plummer}, with magnetic wave vector ${\bf Q}\!=\!(\frac{1}{4},\frac{1}{4},0)$ as revealed by early neutron scattering experiments \cite{mathieu}. Since pure Sr$_3$Ru$_2$O$_7$ does not show any long-range magnetic order, it is clear that the latter is induced by the $S\!=\!2$, 3$d$-Mn$^{3+}$ impurities \cite{mathieu,hossain_PRL}; we emphasize however that the ordering -- and particularly the ${\bf Q}\!=\!(\frac{1}{4},\frac{1}{4},0)$ magnetic wave vector -- are independent of the precise 5-10\,\% Mn concentration, indicating that the role of Mn is merely that of triggering and/or stabilizing a tendency already incipient in the parent compound Sr$_3$Ru$_2$O$_7$ \cite{hossain_PRB}.\\

\noindent
{\bf Resonant scattering at the Ru L-edge.} We have performed REXS experiments on Sr$_3$(Ru$_{1-x}$Mn$_x$)$_2$O$_7$ on both Ru and Mn $L_{2,3}$ edges, for a range of Mn concentrations. On a 10\% Mn-substituted sample at 20\,K, REXS at the Ru $L$-edge detects a forbidden superlattice diffraction peak appearing at ${\bf Q}=(\frac{1}{4},\frac{1}{4},0)$. The Ru L edge measurements provide a critical information: while the ${\bf Q}\!=\!(\frac{1}{4},\frac{1}{4},0)$ modulation observed by neutron scattering points toward an electronic modulation or order in the system \cite{mathieu,Plummer}, one needs to exclude that this is not merely associated with a Mn clustered phase or a structural ordering of the Mn impurities, in which case the host Ru valence electrons would not participate. As revealed by the photon-energy dependence of the $(\frac{1}{4},\frac{1}{4},0)$ peak in Fig.\,\ref{Fig3}(a), the intensity enhancement observed at both $L_{2,3}$ Ru absorption edges -- with onset at $T_{order}$ of the AF phase as shown in Fig.\,\ref{Fig3}(b) -- demonstrates that the new order is electronic and not chemical/structural, and pertains to the system as a whole. In addition, the magnetic character of the order is confirmed by the azimuthal dependence of the order parameter in REXS experiments \cite{hossain_PRB}. This sets the stage to investigate the possibility of studying the magnetic superstructure by REXS measurements performed at the Mn-impurity L edge. \\
\begin{figure}[t]
\centerline{\epsfig{figure=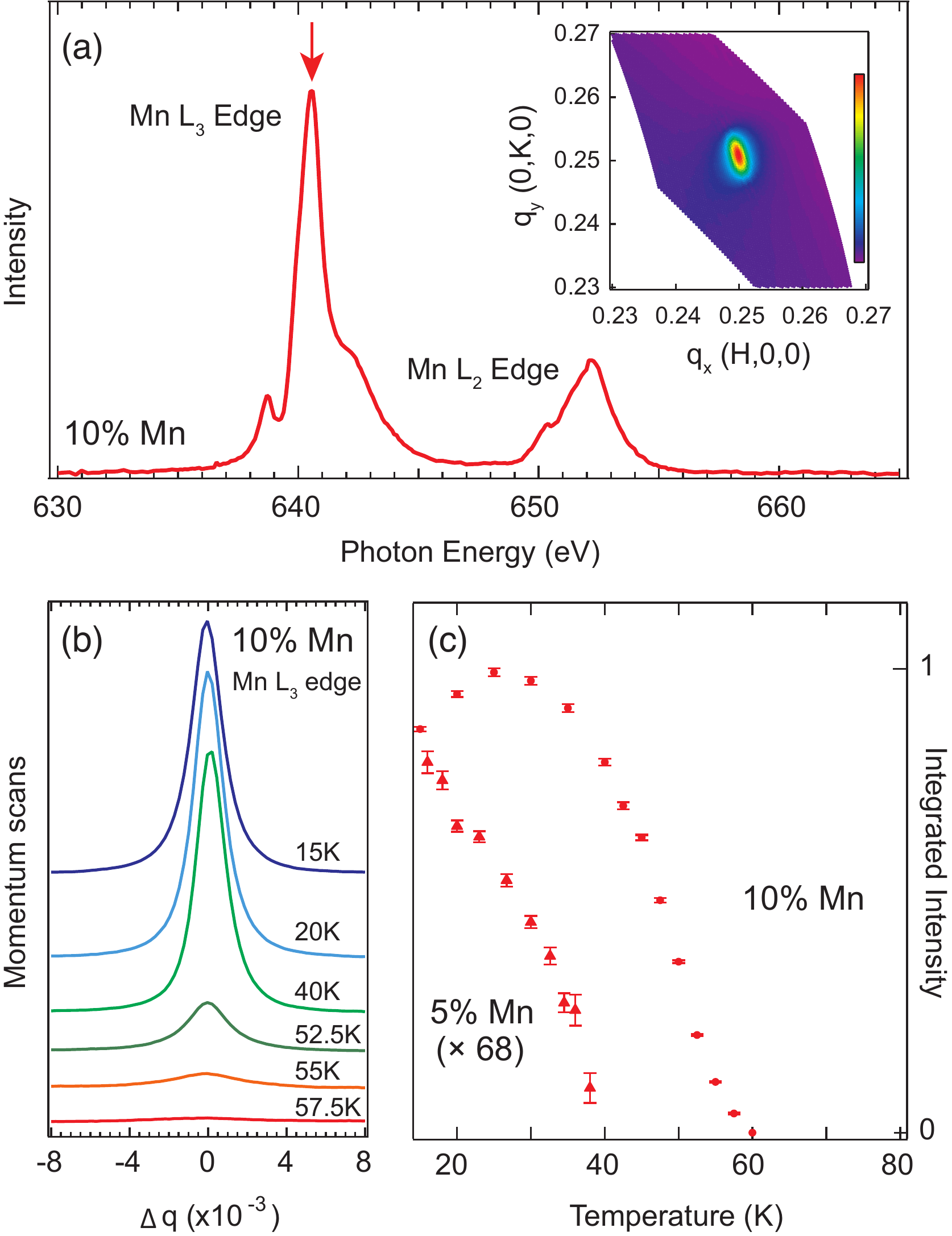,width=1.0\linewidth,clip=}} 
\caption{(a) Mn resonance profile for the $(\frac{1}{4},\frac{1}{4},0)$ superlattice diffraction peak measured at 20\,K on Sr$_3$(Ru$_{1-x}$Mn$_x$)$_2$O$_7$ with $x\!=\!0.1$ (the arrow at 641\,eV indicates the energy used in the Mn-edge REXS experiments). Inset: full reciprocal space map of the superlattice peak. (b) Temperature dependence of the Mn $L_{3}$ edge $(\frac{1}{4}+\Delta q,\frac{1}{4}+\Delta q,0)$ momentum scans for $x\!=\!0.1$. (c) Temperature dependence of the integrated intensity of the $(\frac{1}{4},\frac{1}{4},0)$ peak for 5\% and 10\% Mn-substitution.}
\label{Fig4}
\end{figure}
\begin{figure*}[t!]
\centerline{\epsfig{figure=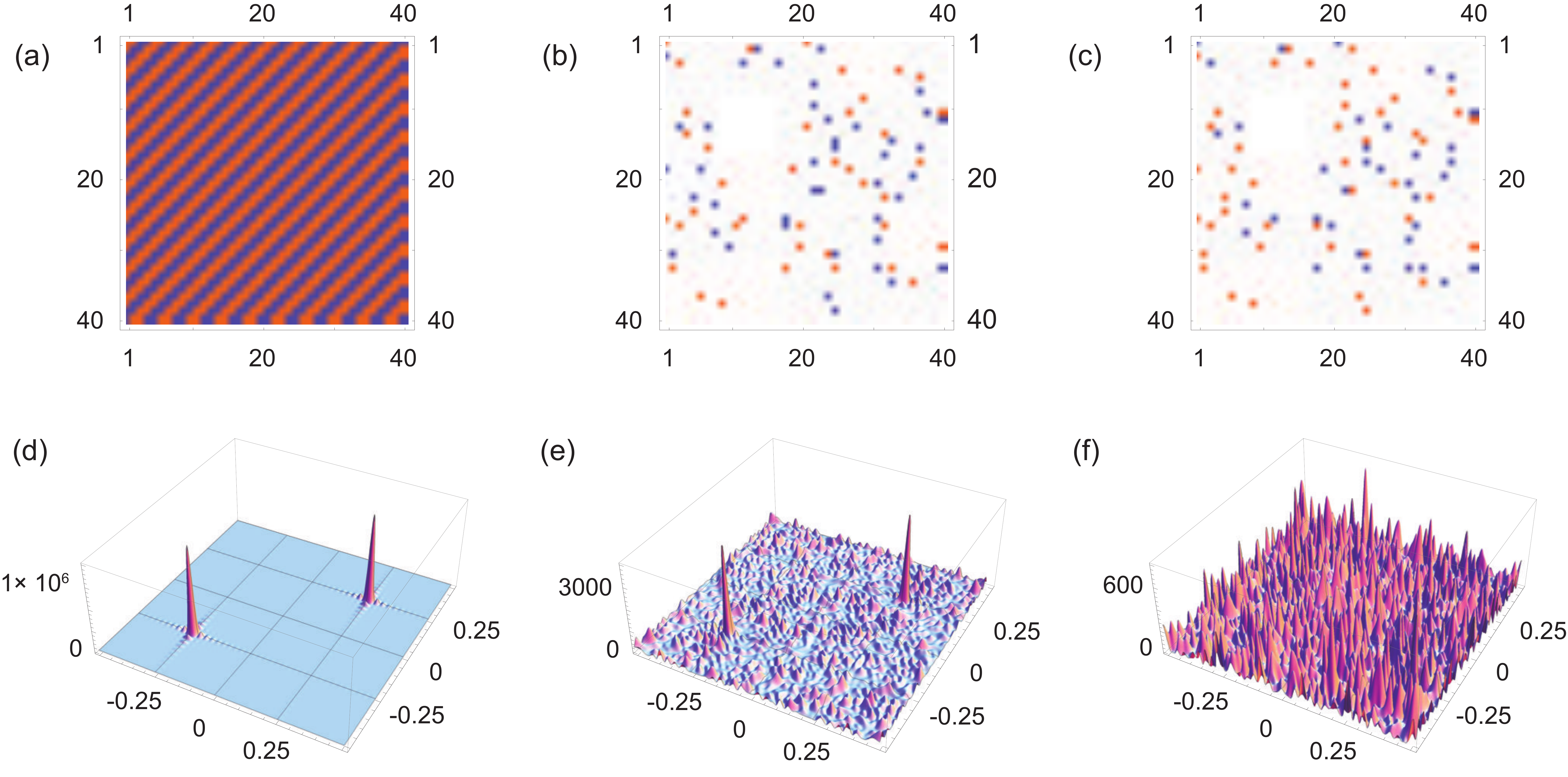,width=1.0\linewidth,clip=}} 
\caption{(a) Square lattice ($40\!\times\!40$ sites) with an up-up-down-down zig-zag spin order (red/up, blue/down). (b) Randomly selected 5\% lattice sites from panel (a), with the same spin correlations as in (a). (c) Same 5\% lattice sites as in (b), but now with random spin orientation for the selected sites. (d,e,f) Reciprocal space map of the ${\bf Q}= (\frac{1}{4},\frac{1}{4},0)$ diffraction peaks.}\label{Fig2}
\end{figure*}

\noindent
{\bf Resonant scattering at the Mn L-edge.}  At the Mn $L_{2,3}$-edge, REXS measurements were performed on 5 and 10\% Mn-substituted Sr$_3$Ru$_2$O$_7$ and indeed exhibit superlattice diffraction peak. For clarity, only the data from 10\% Mn substitution are shown in Fig \,\ref{Fig4}(a), which presents the energy dependence of the ${\bf Q}=(\frac{1}{4},\frac{1}{4},0)$ superlattice peak intensity at the Mn $L_{2,3}$ resonance profile at $T\!=\!20$\,K (its full 2-dimensional reciprocal space map is shown in the inset). Also in the Mn case the forbidden superstructure reflections are detected only below $T_{order}$; and as for the Ru $L$-edge measurements, the magnetic character of the Mn $L$-edge superlattice peak has again been confirmed by the azimuthal dependence of the order parameter \cite{hossain_PRB}. The magnetic order progressively gain strength upon reducing temperature and increasing Mn concentration. This is shown as a function of temperature in Fig.\,\ref{Fig4}(b) based on the raw data for the $(\frac{1}{4}+\Delta q,\frac{1}{4}+\Delta q,0)$ momentum scans measured on the 10\% Mn sample at 641\,eV ($L_{3}$-edge). A summary for both 5 and and 10\,\% Mn substituted samples is presented in Fig.\,\ref{Fig4}(c), where the integrated peak intensities of the longitudinal momentum scans $(\frac{1}{4}+\Delta q,\frac{1}{4}+\Delta q,0)$ are plotted versus temperature. Following the experimental evidence given above, in the remainder of the paper we will demonstrate -- both analytically and computationally -- how resonant scattering at the impurity edge can be observed, and what deeper insights might this provide.\\

\noindent
{\bf Scattering from random impurities.} To understand how it is possible to observe superlattice diffraction signal from random impurities, we can start from the idealized case of 100\% Mn substitution. Fig.\,\ref{Fig2} presents such MnO$_2$ square lattice with an up-up-down-down zig-zag spin order (red/up, blue/down). This magnetic pattern satisfies the symmetries obtained in the azimuthal dependence measured by REXS in Ref.\,\onlinecite{hossain_PRB}. For the 100\% spin-spin correlations of Fig.\,\ref{Fig2}(a), the corresponding reciprocal space map is characterized by well-defined (i.e. infinitely sharp for an infinite lattice) magnetic superstructure peaks ${\bf Q}\!=\!(\frac{1}{4},\frac{1}{4},0)$ and $(-\frac{1}{4},-\frac{1}{4},0)$, as shown in Fig.\,\ref{Fig2}(d). When 5\% of the sites are selected randomly from Fig.\,\ref{Fig2}(a) while retaining the very same spin-spin correlations as in Fig.\,\ref{Fig2}(b) -- mimicking the 5\% Ru-Mn substitution in the RuO$_2$ plane -- equally sharp magnetic peaks (with reduced intensity) are still observed [Fig.\,\ref{Fig2}(e)]. On the contrary, when the very same subset of lattice sites is given a random spin orientation as in Fig.\,\ref{Fig2}(c), superlattice peaks can no longer be observed in the reciprocal space map [Fig.\,\ref{Fig2}(f)]. This is to be expected since the sites are random and -- more importantly -- spin-spin correlations are lost. The somewhat unintuitive result in Fig.\,\ref{Fig2}(e) can be explained as follows: although the 5\% Mn atoms are randomly distributed, as long as they reside at sites belonging to the host square lattice and retain proper phase relation within the spin ordering, they will scatter coherently giving rise to the same magnetic Bragg peaks and even reflecting the same correlation length.

The nonvanishing scattering signal from random Mn impurities obtained in the simulations of Fig.\,\ref{Fig2} can also be demonstrated analytically based on the formal expression of the scattering amplitude \cite{book}:
\begin{eqnarray}
A({\bf Q}) && = \sum^{all \ sites}_{{\bf R}_n}f_{{\bf R}_n} e^{i {\bf Q} \cdot {\bf R_n}}\times\delta_{{\bf R}_n, {\bf R}_{Mn}} \nonumber \\
&& = \sum^{Mn \ sites}_{{\bf R}_{Mn}} f_{{\bf R}_{Mn}} e^{i {\bf Q} \cdot {\bf R}_{Mn}} \, ,\nonumber 
\end{eqnarray}
\begin{figure}[b!]
\centerline{\epsfig{figure=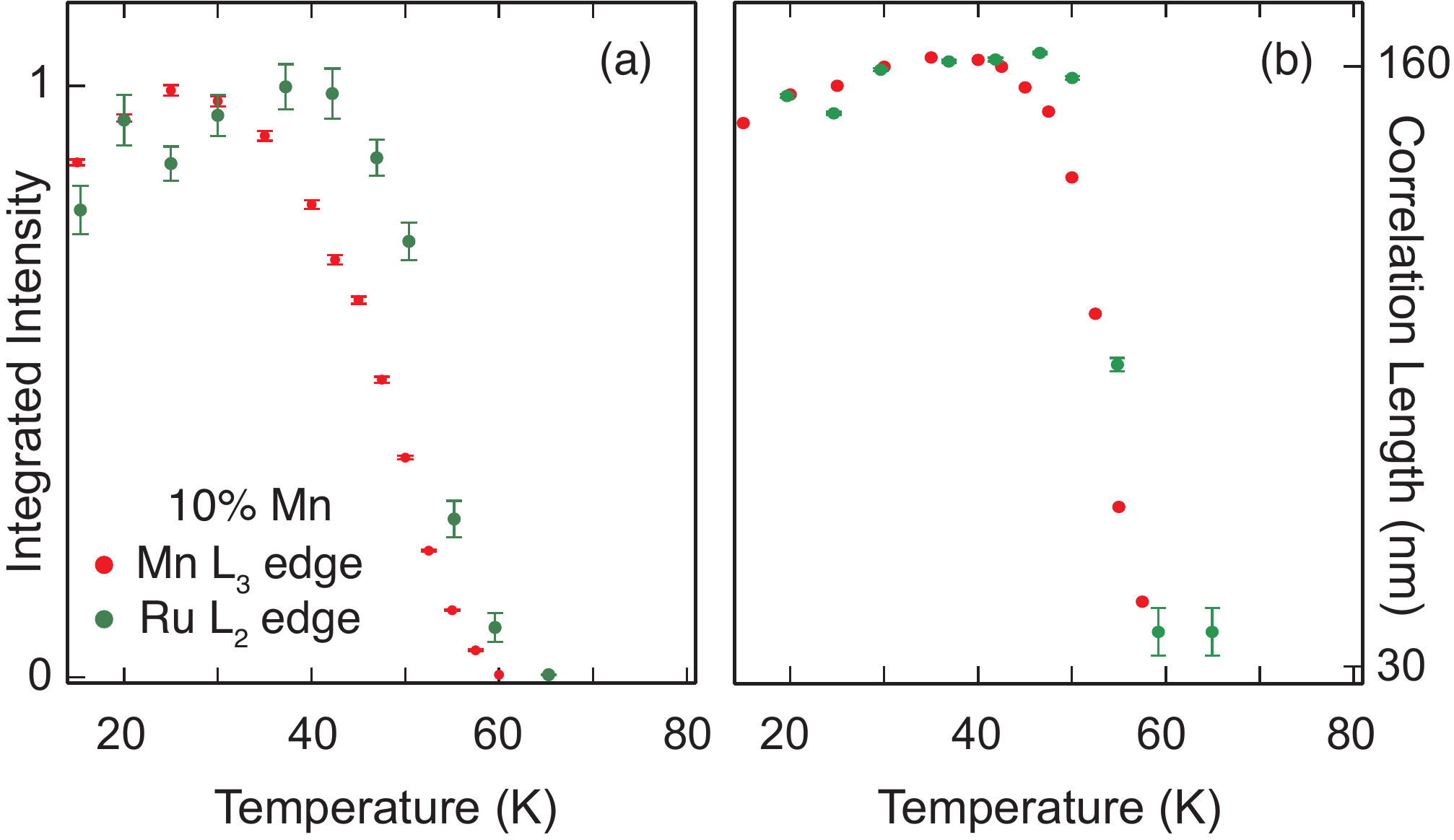,width=1.0\linewidth,clip=}}
\caption{Temperature dependence of the (a) normalized integrated intensity and (b) correlation length of the $(\frac{1}{4},\frac{1}{4},0)$ superlattice order, measured at the Mn (red) and Ru (green) L-edges for 10\% Mn-substitution.
}\label{Fig5}
\end{figure}
\begin{figure*}[t!]
\centerline{\epsfig{figure=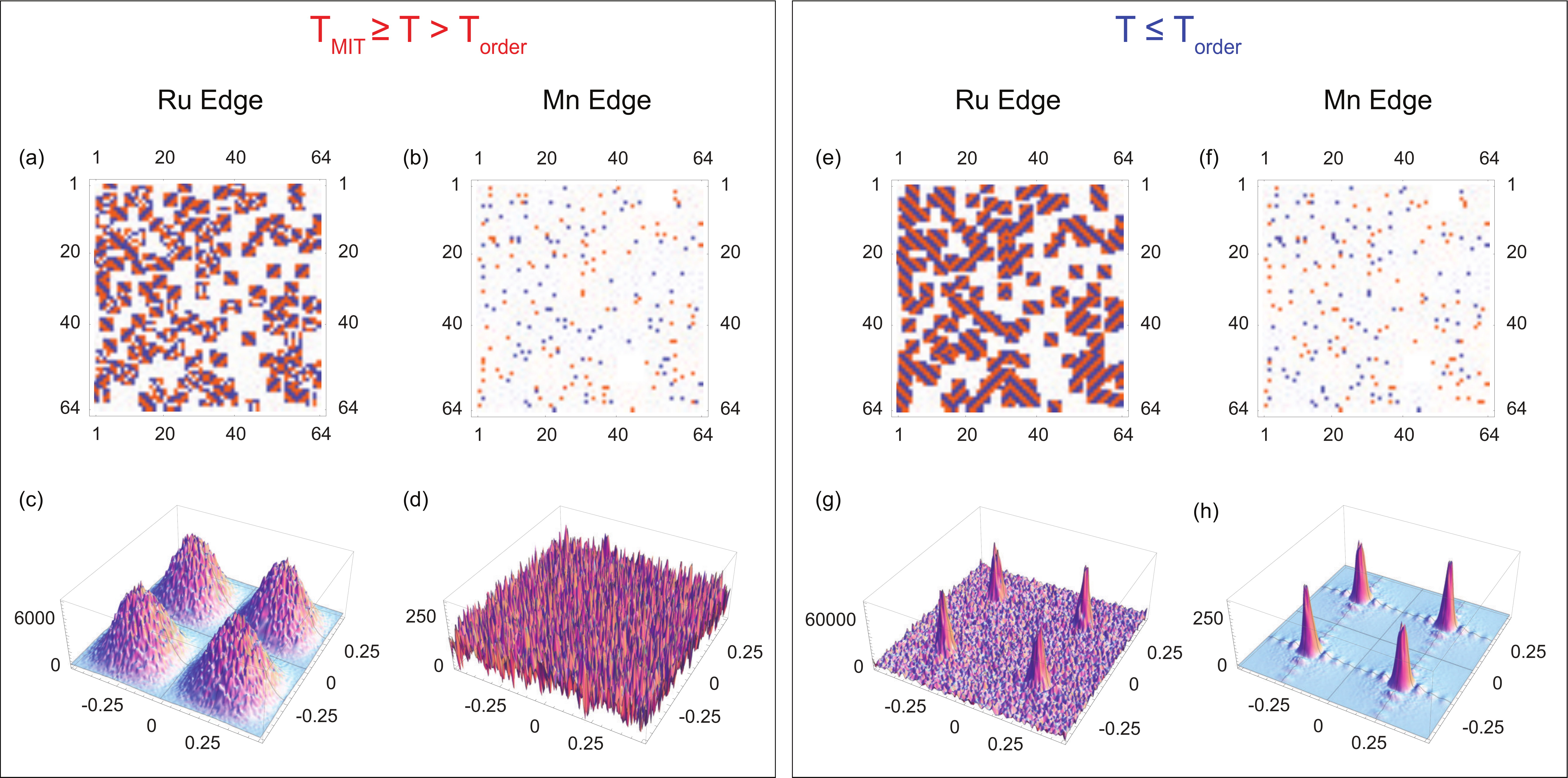,width=1.0\linewidth,clip=}}
\caption{(a) Zig-zag antiferromagnetic islands (red/spin-up, blue/spin-down) around Mn impurities, just below the metal-insulator transition temperature $T_{MIT}$, i.e. the percolation threshold before the onset of long-range order -- the islands are not correlated with one-another yet. (b) Randomly selected 5\% lattice sites from panel (a), representing the 5\% Mn impurities, with the same spin correlations. (e,f) Reciprocal space map of the ${\bf Q}= (\pm\!\frac{1}{4},\pm\!\frac{1}{4},0)$ diffraction peaks for Ru and Mn edges, calculated from the corresponding top panels. (c,d,g,h) Same as (a,b,e,f), now below the long-range antiferromagnetic ordering temperature $T_{order}$, with partial magnetic correlations having been established between different antiferromagnetic islands.}\label{Fig6}
\end{figure*}

where ${\bf R}_n$ denotes the atomic positions, $f_{{\bf R}_{n}}$ is the atomic form factor, and $\delta_{{\bf R}_n, {\bf R}_{Mn}}$ is 1 at the Mn impurity sites and zero elsewhere. It is clear from the above expression that while $\delta_{{\bf R}_n, {\bf R}_{Mn}}$ removes all the non-Mn sites, the phase relation between the Mn sites -- encapsulated in the exponential term -- remains the same; hence, the signal from these sites interferes constructively for the same {\bf Q} superlattice vector. It follows that the scattering intensity from random impurities can be nonzero and is proportional to the square of the number $N_{Mn}$ of Mn atoms, $I({\bf Q})=|A({\bf Q})|^2 \propto N^2_{Mn}$. This demonstrates that by reducing the number of Mn impurities in Sr$_3$(Ru$_{1-x}$Mn$_x$)$_2$O$_7$ one only loses signal proportionally to the reduced occupation of the Mn sublattice, and the rest of the Mn atoms still contribute to the diffraction superstructure as long as they have the same ${\bf Q}= (\frac{1}{4},\frac{1}{4},0)$ and $(-\frac{1}{4},-\frac{1}{4},0)$ phase relationship. Since the structure factor is the Fourier transform of the two-point correlation function \cite{lubensky}, removing Mn atoms also does not change the correlation length.\\

\noindent
{\bf Comparison between Ru and Mn edge.} Further insights can be obtained from the direct comparison of the temperature dependence of Ru- and Mn-edge magnetic peak intensity, and corresponding correlation length (defined as $2\pi/{\rm FWHM}$, where FWHM is the full-width half-maximum of the momentum scans). This is presented in Fig.\,\ref{Fig5}(a,b) for the 10\% Mn-substituted sample. Fig.\,\ref{Fig5}(a) shows an intensity onset $\sim\!5$\,K higher on Ru than Mn (i.e., 65 versus 60\,K), as well as a different shape of the temperature dependence (with a steeper increase at the Ru edge). This is accompanied by the smooth increase of Ru- and Mn-edge correlation lengths upon reducing the temperature [Fig.\,\ref{Fig5}(b)]. Overall, the Mn order parameter is lagging behind the Ru one. The further decrease of the correlation length below 30\,K has also been seen in reentrant spin glasses, due to the interplay of competing interactions and disorder \cite{Maletta,Aeppli}). It is important to note that the order is never truly long-range at any temperature, with the highest value for the intrinsic correlation length being only 160\,nm, as determined at both Ru and Mn $L$ edges; in addition, as shown in Ref.\,\onlinecite{hossain_PRB}, the correlation length is even shorter for lower doping levels.

Overall, these data indicate that the antiferromagnetic correlations first appear in the RuO$_2$ islands surrounding the Mn impurities; however, since initially these islands do not overlap, the Mn impurity spins are not correlated, and a somewhat incoherent magnetic signal is observed only at the Ru edge. When these islands begin to overlap, the RuO-mediated exchange interaction can energetically favor a coherent spin
arrangement between the islands; as a result, the Mn impurities will interfere constructively and long-range order is observed at both edges.  This behavior cannot be explained based on the assumption of a 100\% spin correlation between Mn impurities as in Fig.\,\ref{Fig2}, and should instead be discussed in the context of a percolative phase transition, with an effective percolation threshold achieved at $\lesssim\!5$\% Mn concentration \cite{hossain_PRB}. 

This case is presented in Fig.\,\ref{Fig6}(a), which shows a 64$\times$64 RuO$_2$ lattice in real space, with 5\,\% Mn sites each inducing a 4$\times$4 unit of the magnetic zigzag stripe pattern. As in Fig.\,\ref{Fig2}, blue and red squares in Fig.\,\ref{Fig6}(a) represent up and down spins along the $c$-axis, while the white patches are regions of the RuO$_2$ plane where no magnetism has been induced. Here, however, no correlation between the Mn spins has been imposed ($T_{MIT}\!>T\!>\!T_{oder}$ \cite{hossain_PRB}); the position of the Mn impurities and their corresponding spins are shown in Fig.\,\ref{Fig6}(b). Even though there is no spin correlation between the 4$\times$4 magnetic islands, the reciprocal space map of the scattering intensity from the Ru sites in Fig.\,\ref{Fig6}(e) shows extremely broad and weak scattering peaks stemming from the onset of short-range order below $T_{MIT}$ \cite{hossain_PRB}. Due to the lack of correlation between the Mn spins, no Mn-edge peaks are visible in Fig.\,\ref{Fig6}(f). The situation is very different for correlated Mn spins ($T\!<\!T_{oder}$); this is shown in Fig.\,\ref{Fig6}(c,d), which presents the same sequence of Fig.\,\ref{Fig6}(a,b) but now for an average `magnetically correlated' domain of 68$\times$68\,\AA$^2$ \ ($\sim$16$\times$16 lattice sites), as observed experimentally once long-range order is established below $T_{oder}$ \cite{hossain_PRB}. As shown in Fig.\,\ref{Fig6}(g,h), the superlattice $(\pm\frac{1}{4},\pm\frac{1}{4},0)$ peaks are clearly discernible above the noise -- with a well defined correlation length -- at both Ru and Mn edges.

In addition to demonstrating that impurity-edge superlattice peaks can be measured even in case of a complex percolative phase transition, the results in Fig.\,\ref{Fig6} reveal the origin of the difference in temperature evolution of Ru- and Mn-edge intensity in Fig.\,\ref{Fig5}(a). This can be summarized as follows: the Ru-edge intensity has an earlier onset and also faster rise than that of Mn. Our modelling shows that in the crossover from $T_{MIT}$ to $T_{order}$, Mn impurities begin to get correlated, which defines the onset of the Mn-edge intensity;  however, the number of magnetically correlated sites increases much faster for Ru than Mn, resulting in a much faster REXS intensity rise at the Ru edge. Eventually, in the saturation regime, the correlation length is the same at both edges since this is defined by size of the macroscopically magnetically-ordered domains, which is dependent on the Mn-concentration.

\section*{\large{Discussion}}
This study proves that REXS signal from dilute and randomly distributed impurities can give rise to superlattice diffraction peaks, which can be used to probe the electronic order in an element sensitive manner. Interestingly, the underlying spin order in Sr$_3$(Ru$_{1-x}$Mn$_x$)$_2$O$_7$ is more difficult to detect at the Ru than at the Mn $L$-edge, where a much larger resonance is observed. In the case of 10\% Mn substitution, as evidenced by the signal-to-noise ratio and corresponding error bars in Fig.\,\ref{Fig4}(c), the scattering signal from Ru is already much weaker than that of Mn, although Ru occupies 90\% of the lattice sites. Once Mn substitution is reduced to 5\%, the Ru signal is swamped in the background while REXS at the Mn $L$-edge still provides very clean information on the magnetic superstructure. Addressing this additional aspect -- which partly stems from the much stronger magnetic moment and scattering cross section of Mn as compared to Ru -- is beyond the scope of our simulations in Fig.\,\ref{Fig6}; most importantly, however, this also shows the promise of using a suitable dilute impurity -- with large scattering cross-section, magnetic-moment, or orbital polarizability -- as a ``marker'' or ``magnifying glass" to detect weak spin/charge/orbital order in a host system that is difficult to probe via the majority lattice elements. 

Beyond cross section considerations, we note that in many transition-metal oxides the electronic-ordering superlattice constant is too small to be detectable at the strong L$_{2,3}$ resonances, because of the too long photon wavelength at those energies. Substitution with an element of the same valence and similar size -- but with a much higher absorption-edge resonance energy, corresponding to much shorter wavelength photons -- could be used to gain access to the electronic ordering, enabling the study of the associated phases and phase transitions. If in addition one can scatter from both impurity and host atoms selectively, this approach would become extremely powerful in the study of spatially inhomogeneous orders in bulk, surfaces, interfaces, and heterostructures. At an interface or heterostructure, new electronic states can occur due to interaction between interfacial atoms; the role of electronic order and interaction across the interface might be clarified by element selective and/or impurity-edge REXS performed right at the interfacial region.

Finally, these ideas and approaches can also be generalized to other techniques, such as e.g. resonant inelastic x-ray scattering (RIXS) and time-resolved resonant x-ray scattering. With the development of brighter x-ray sources and higher resolution momentum resolved capabilities, one can take advantage of impurity edge RIXS experiments to study momentum-resolved elementary excitations (phonons, magnons, orbitons), involving host and impurity degrees of freedom and also theoretical predictions on subtle issues like whether noninteracting atoms contribute coherently to the inelastic scattering cross section \cite{ma}. In addition, with the availability of coherent two-color time-resolved probes at the next generation free-electron laser sources, such as LCLS, one can envision experiments performed simultaneously at the impurity and host edges, to disentangle the causal relationships between lattice, charge, spin, and orbital orders in a variety of complex materials. 

\section*{\large{Methods}}

\noindent
{\bf Light scattering experiments.} REXS measurements were performed at beamlines 8.0.1 at ALS in Berkeley (Mn $L$-edges) and KMC-1 at BESSY in Berlin (Ru $L$-edges). In both cases we used a two-circle ultra-high-vacuum diffractometer in horizontal scattering geometry, with the incident photon beam polarized parallel to the diffraction plane ($\pi$). The scattered signal contained polarization components both parallel ($\pi'$) and perpendicular ($\sigma'$) to the diffraction plane. The correlation lengths in Fig.\,5 are defined as $2\pi/{\rm FWHM}$, where FWHM is the full-width half-maximum of the Lorentzian fit to the momentum scans (with $x$-axis converted to \AA$^{-1}$ units). At the Mn and Ru $L$ edges, our experimental angular resolutions are $\lesssim1^\circ$ and $0.4^\circ$, respectively.\\

\noindent
{\bf Sample preparation.} Sr$_3$(Ru$_{1-x}$Mn$_x$)$_2$O$_7$ single crystals grown by the floating zone technique \cite{mathieu} were cut and polished along the (110) direction. The samples were mounted on cryogenic manipulators, which allow a polar ($\theta$) and azimuthal ($\phi$) angle rotation of the sample about the scattering vector, in the temperature range 20-300\,K. Note that the diffraction peaks are indexed with respect to the undistorted tetragonal $I$4$/mmm$ unit cell with axes along the RuO bond directions ($a_0\!=\!b_0\!\simeq\!3.9$\,\AA).

\bibliographystyle{plain}

\subsection{Acknowledgements}
We thank M.Z. Hasan for the use of the ALS scattering chamber and E. Schierle, E. Weschke for the BESSY/HMI scattering chamber and M. Le Tacon for fruitful discussions. This work was supported by the Max Planck - UBC Centre for Quantum Materials, the Killam, Alfred P. Sloan, Alexander von Humboldt, and NSERC's Steacie Fellowships (A.D.), the Canada Research Chairs Program (A.D. and G.A.S.), ALS and NSERC postdoctoral fellowships (M.A.H), NSERC, CFI, CIFAR Quantum Materials, and BCSI. H.-H.W. and C.S.-L. are supported by DGF through SFB 608. The Advanced Light Source (ALS) is supported by the Director, Office of Science, Office of Basic Energy Sciences, of the U.S. Department of Energy under Contract No. DE-AC02-05CH11231.

\subsection{Author Contributions}
M.A.H., Y.-D.C., B.K., and A.D. planned the experiments. M.A.H., Y.-D.C., A.G.C.G. carried out the Mn L edge soft x-ray diffraction experiments at ALS and M.A.H., I.Z., B.B., H.-H.W., and C.S.-L. carried out the Ru L edge experiments at BESSY. M.A.H. and B.B. analyzed the Mn and Ru edge data, respectively. M.A.H., Y.-D.C., J.G., B.B., D.G.H., J.D.D., Z.H., B.K., G.A.S., and A.D. contributed to the data interpretation and modelling discussions. M.A.H. wrote the modelling codes presented in Fig.\,\ref{Fig2} and \ref{Fig6}. R.M., Y.T., S.S., H.T., and Y.Y. grew the samples. M.A.H. and A.D. wrote the manuscript with contributions from the other authors. A.D. is responsible for overall project direction, planning, and management.

\subsection{Additional Information}
The authors declare to have no competing financial interests, or any other interests that might be perceived to influence the results and/or discussion reported in this article.
\end{document}